# The latest on Apertif

**Tom Oosterloo[1,a,b], Marc Verheijen[b] and Wim van Cappellen[a]**
[a]*Astron - Netherland Institite for Radio Astronomy*
*Postbus 2, 7900 AA Dwingeloo, The Netherlands*
[b]*Kapteyn Astronomical Institute, University of Groningen*
*Postbus 800, 9700 AV Groningen, The Netherlands*
*E-mail:* `oosterloo@astron.nl`

We describe a Phased Array Feed (PAF) system, called Apertif, which will be installed in the Westerbork Synthesis Radio Telescope (WSRT). The aim of Apertif is, at frequencies from 1.0 to 1.7 GHz, to increase the instantaneous field of view of the WSRT to 8 deg$^2$ and its observing bandwidth to 300 MHz with high spectral resolution. This system will turn the WSRT into an effective survey telescope with scientific applications ranging from deep surveys of the northern sky of HI and OH emission and polarised continuum to efficient searches for pulsars and transients. We present results obtained with a prototype PAF installed in one of the WSRT dishes. These results demonstrate that at decimetre wavelengths PAFs have excellent performance and that even for a single beam on the sky, they can outperform single feed radio dishes. PAFs turn radio telescopes into very effective survey instruments. Apertif is now fully funded and the community invited to express their interest in using Apertif (see http://www.astron.nl/radio-observatory/call-expressions-interest-apertif-surveys)

---

[1] Speaker

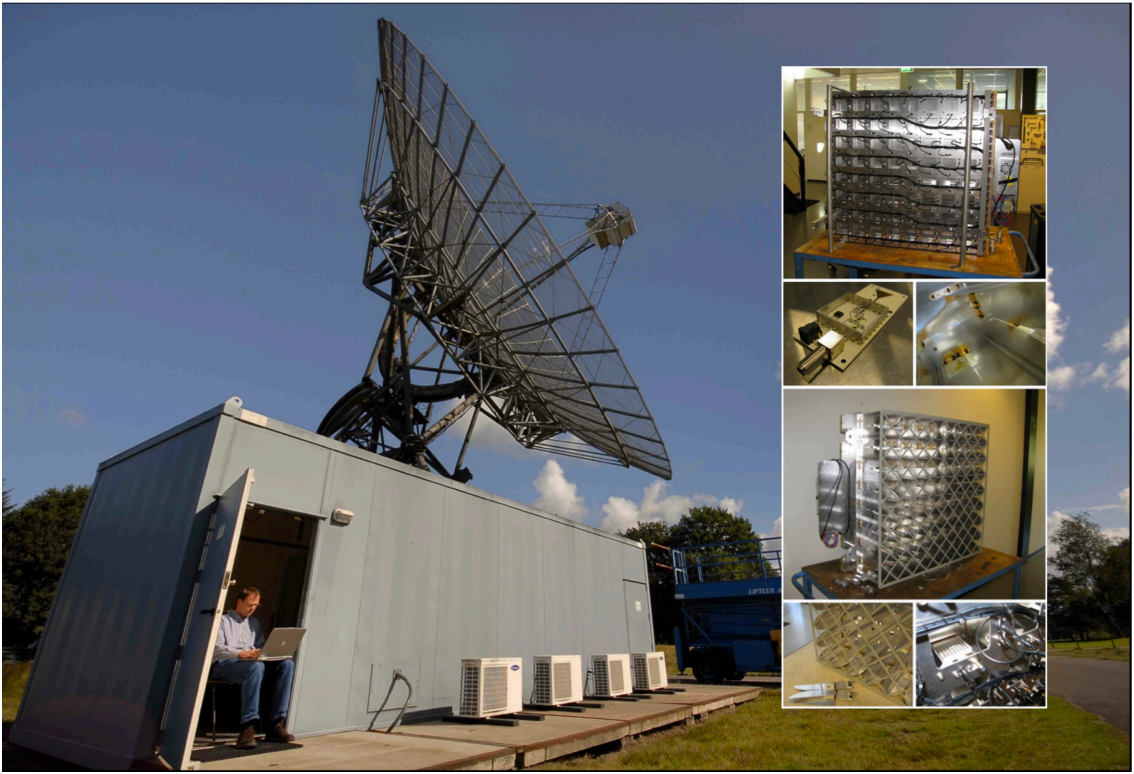

**Figure 1.** The WSRT telescope RT5 equipped with the APERTIF prototype. Shown in the inserts is the Vivaldi array in the lab

## 1. Introduction

Radio astronomy is a very successful branch of astronomy which has made several fundamental contributions to science, for example, the detection of the cosmic microwave background, the discovery of pulsars and detailed spectroscopic studies of the cold interstellar medium in the Milky Way and other galaxies. However, in order to keep the momentum of discovery, the performance of radio instruments has to be improved continuously. One of the main limitations of many current radio telescopes is that they are poor survey instruments. Because of their limited field of view, it is very expensive in terms of observing time to make deep observations of large volumes of space in order to detect sufficiently large samples of faint sources. Such deep, wide-field studies are called for in the context of many very relevant astronomical topics, such as the role of gas in the evolution of galaxies and the properties of the transient sky.

The obvious solution is to replace the single-feed systems currently employed in many radio telescopes with arrays of detectors. This effectively turns a radio dish into a radio camera. On single-dish radio telescopes this has been done by installing multiple feeds in the focal plane which provide several beams on the sky. Examples are the successful Parkes multi-beam system [6,7], the Arecibo multi-beam system AlfAlfa [3] and the Effelsberg Bonn HI Survey [5].

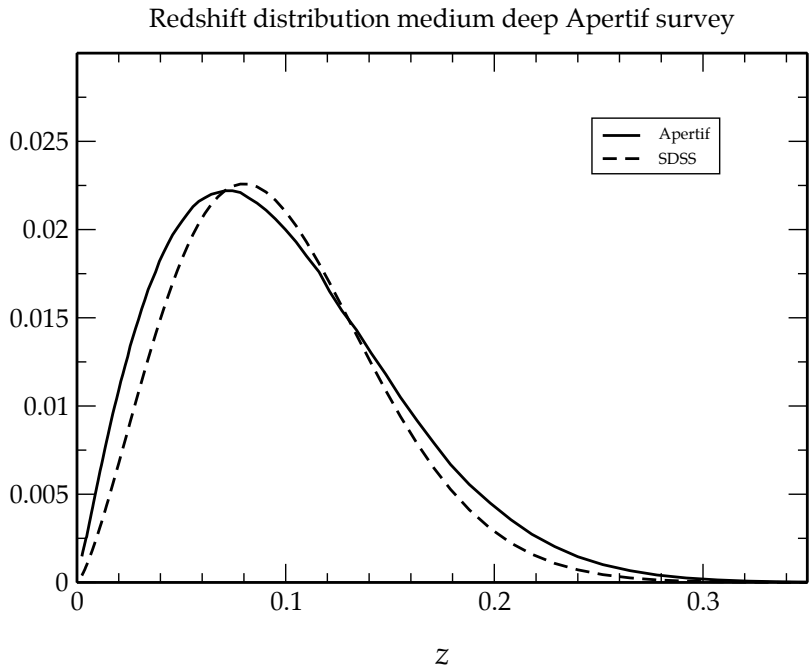

**Figure 2**. Redshift distribution of HI detections in a medium-deep HI survey with Apertif. A large fraction of the detections will be at $z > 0.1$. For comparison, the redshift distribution of the spectroscopic observations of the Sloan survey (SDSS) is also given, showing that there is a very good match between the two datasets

Current technology offers a different method to form multiple beams on the sky and to do that in a more efficient way. The basic idea is to employ phased-array technology in the focal plane of a radio telescope to form multiple beams on the sky. The advantage of this technique is that the radiation field in the focal plane is *fully* sampled. This offers much greater flexibility in forming many beams and allows, at the same time, to optimise the performance of the telescope dish for maximum gain or other properties. Apertif (``APERture Tile In Focus") is such a system and is being developed for the Westerbork Synthesis Radio Telescope (WSRT). The aim of Apertif is to turn the WSRT into an instrument that will perform deep northern-sky imaging surveys in spectral line and polarised continuum at frequencies between 1 and 1.7 GHz. In this paper we briefly describe the science case of Apertif, we discuss the technical details of this focal plane array (PAF) system and report on results obtained with a prototype PAF installed in one of the WSRT dishes.

## 2. Wide-field radio astronomy at GHz frequencies

### 2.1 Spectral line studies

One of the main scientific applications of wide-field radio telescopes operating at GHz frequencies is to observe large volumes of space in order to make an inventory of the neutral hydrogen in the Universe. With such information, the properties of the neutral hydrogen in galaxies as function of mass, type and environment can be studied in great detail, and, importantly, for the first time the *evolution* of these properties with redshift can be addressed. Given the current sensitivity and the limited field of view of radio telescopes, statistical studies of the neutral hydrogen in galaxies are either limited to relatively small volumes or to bright sources. Hence, the number of faint and/or distant objects detected is relatively small and, in particular, the issue of evolution cannot be studied. The large field of view offered by radio telescopes with PAFs makes it possible to observe much larger volumes at high sensitivity. The improvement will be dramatic. Currently, we know about the HI in about 20,000-30,000 galaxies, of which a few hundred are at redshift above 0.1. Moreover, the large majority of these data consists of single-dish measurements, so we only know about the global HI properties. Surveys done with instruments like Apertif will result in a few hundred thousand detections, most of which will be above *z* = 0.1 (see Figure 2). Moreover, the spatial resolution of Apertif data is such that most detections will be spatially resolved. This huge improvement will bring extragalactic HI studies to a different level and makes it possible to address several important questions. One of these is the role gas and gas accretion plays in the evolution of galaxies. Many observational studies have shown that in the last 7 Gyr, the star formation rate density has declined with about a factor 10. This strong evolution is not reflected in the neutral hydrogen because the overall cosmological density of neutral hydrogen shows a much more modest evolution with redshift. This is in contrast with the fact that, given their star formation rate and their gas content, the timescales on which gas in galaxies is consumed by star formation typically are only a few Gyr. Hence, one would expect to see a connection between the evolution of star formation rate and neutral hydrogen content of galaxies. Clearly, something is missing in our understanding of how galaxies evolve. Instruments like Apertif will be able to survey significant volumes for neutral hydrogen out to large distances. It is particularly interesting if these HI surveys are done in areas where also good optical information is available, such as the Sloan area (Figure 2). Additionally, a large number of OH megamasers will be detected in such a survey. An Apertif HI survey will allow to directly observe the evolution of the neutral hydrogen content of galaxies and how this evolution differs in different environments and for different galaxy masses and types.

Another important application of wide-field HI surveys is the study of the smallest galaxies in the Local Volume. Current studies of the faint end of the HI mass function have been able to determine the statistics of galaxies with HI masses down to about $10^6 \, M_\odot$. It is of great interest to push this to lower HI masses, because this is the mass range where galaxies may become too small to form stars. The smaller a galaxy is, the less efficient it becomes in converting its gas into stars. One may therefore expect to find a large population of very small galaxies that are, relatively to their stellar content, very gas rich. On the other hand, given the small masses and the associated very shallow gravitational potentials, it is also possible that, if some star formation occurs, the stars and the supernovae will blow out the gas from such small galaxies. Therefore, perhaps the smallest galaxies are gas poor. In addition, these small galaxies are also very vulnerable to destructive effects caused by nearby large galaxies, so an environmental dependence is expected. The only way to establish how (un)common such small galaxies are is using wide-field survey. Given their small HI mass, such objects can be detected, in reasonable integration times, only out to a few Mpc. Therefore, the optimum way to build up a large enough volume so to detect these galaxies in sufficient numbers is by observing large areas on the sky.

**2.2 Radio continuum and polarisation**

Because of the wide observing band of 300 MHz that will be used by Apertif, the spectral line surveys will also produce, from the same observations, very sensitive continuum surveys.

|  | **WSRT** | **Apertif** |
|---|---|---|
| # receiving elements | 1 horn (full pol) | 121 Vivaldi (full pol) |
| # beams on sky | 1 | 37 |
| field of view | 0.3 deg$^2$ | 8 deg$^2$ |
| frequency range | 115 - 8650 MHz | 1000 - 1750 MHz |
| $T_{sys}$ | 30-35 K | 50-55 K |
| aperture efficiency | 55% | 75% |
| bandwidth | 160 MHz | 300 MHz |
| # channels | 1024 | 16384 |

**Table 1:** Comparison of the main instrumental parameters of the WSRT and of Apertif

The expected noise of the continuum images is, depending on survey strategy, 5-20 μJy beam$^{-1}$. Hence, Apertif will produce continuum surveys that are 40-100 times deeper than the NVSS. This will allow to study normal star forming galaxies, in continuum, out to *z* = 1, and systems with more intense star formation out to larger distances, as well as the evolution of weak AGN. An interesting synergy exists with the continuum surveys planned to be done with Lofar. The relative sensitivities of Apertif and Lofar are such that in the planned surveys, a source with spectral index of -0.7 will be detected with both instruments at the same significance level. Therefore, a very similar population of sources will be detected. On the other hand, the combination of the two instruments can be used to find interesting objects, such as high-redshift objects with a steep spectral index that will be detected by Lofar, but not by Apertif.

The broad observing band makes it also possible to use Rotation Measure (RM) synthesis on polarised sources to detect the foreground electro-magnetic medium. Using this technique, a grid of RM values will be observed over the sky with much denser sampling compared to what is available now. Such data will allow to construct very detailed models of the Galactic magnetic field.

**2.3 Pulsars and Transients**

Compared to other SKA pathfinders, Apertif is an effective instrument for finding pulsars. In general, to use an interferometer for pulsar searches, an area on the sky corresponding to the field of view of a single element has to be tiled with tied-array beams which have a size corresponding to the size of the array. For large, sparse arrays, this requires many tied-array beams and, hence, very large resources. However, due to the east-west layout of the WSRT, the instantaneous tied-array beams are one-dimensional fan beams. Therefore tiling the field of view is only a 1-dimensional issue and a relatively small number of tied-array beams is needed to tile the total field of view. The trick is, by observing a field at different hour angles, different sets of fan beams are formed, each with a different orientation on the sky. Pulsars will then be detected at the intersection of these differently oriented fan beams. Using models of the Galactic pulsar population, the expectation is that Apertif may find more than 1,000 new pulsars.

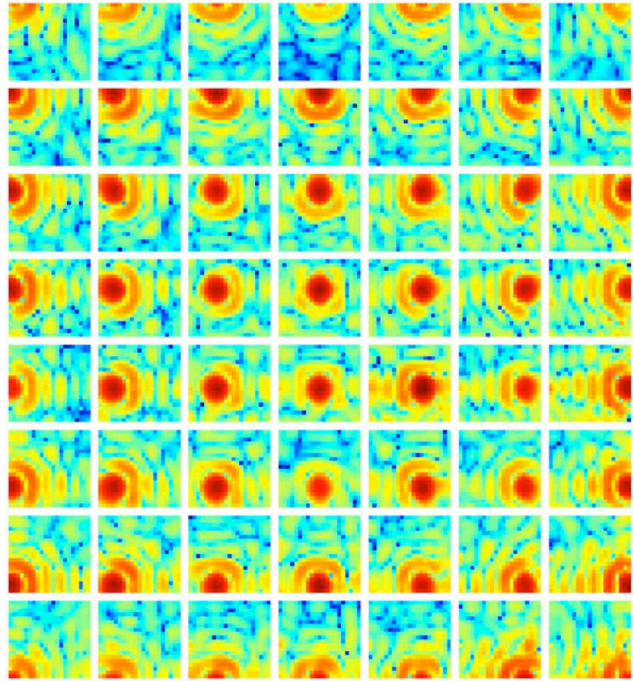

**Figure 3**. Response patterns on the sky of the individual Vivaldi elements of the PAF. Each panel covers the same 3° x 3° on the sky and shows the primary beam of a given element.

For transient radio sources, the new possibilities are also very exciting. A large discovery space will be opened up by new instruments like Lofar, ASKAP and Apertif. One possibility is to observe regions on a regular basis in order to pick up transients. This also offers the possibility to study faint variable sources, a population of which little is known. Another possibility is to use Apertif in a ``fly's eye'' mode where each Apertif dish observes a different area on the sky. More than 100 deg$^2$ can then be covered in a single observation.

## 3. The Apertif system and its prototype

The development of the Apertif PAF system builds upon many years of experience with phased arrays at ASTRON, an effort motivated by enabling this technology for the Square Kilometre Array [4,1]. The design specifications for the APERTIF PAF system are that it should operate in the frequency range of 1000 - 1750 MHz, using an instantaneous bandwidth of 300 MHz covered with 16384 channels, a system temperature of 50-55 K and an aperture efficiency of 75 % (see Table 1). For Apertif this amounts to an effective area to system temperature ratio of $A_{\rm eff}/T_{\rm sys}$ ~100 m$^2$ K$^{-1}$, if all 14 telescopes are equipped with PAF. The goal is to have 37 beams on the sky simultaneously, giving an effective field of view (FoV) of 8 deg$^2$. This entire field of view will be imaged with 15 arcseconds spatial resolution over a bandwidth of 300 MHz with a spectral resolution of about 4 km s$^{-1}$. The survey speed of Apertif, and many of the other characteristics, will be very similar to ASKAP, an PAF-equipped telescope which is being constructed in Australia by the Australia Telescope National Facility. This opens the prospect of truly all-sky surveys with uniform properties in line and continuum. Since May 2010 Apertif is fully funded. Apertif will become operational by the end of 2012 and a call for Expressions of Interest for Apertif surveys has been issued (see http://www.astron.nl/radio-observatory/call-expressions-interest-apertif-surveys).

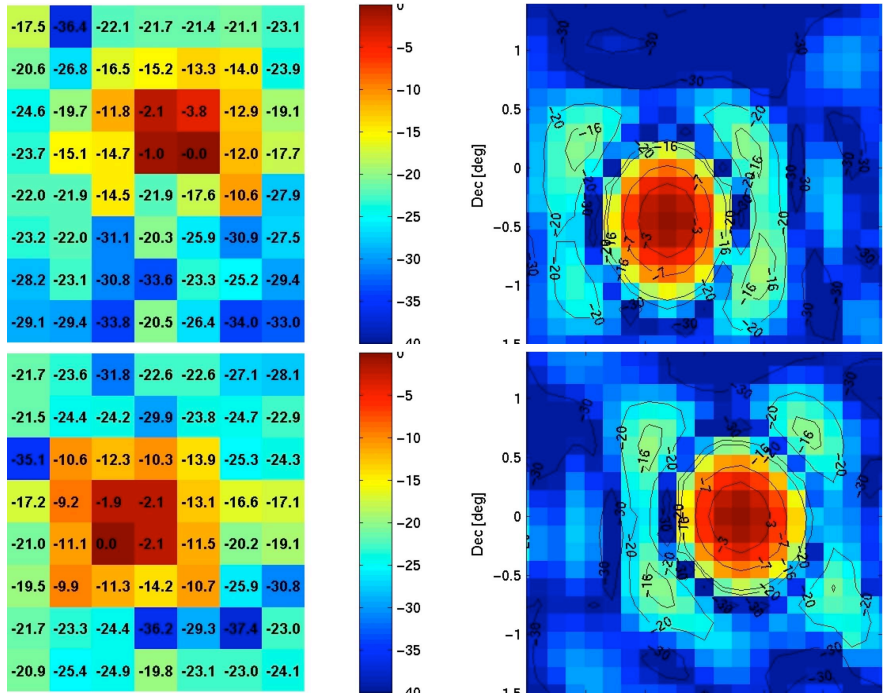

**Figure 4.** Optimised beam patterns (right) and weighting schemes (left) for two directions on the sky. Note the huge improvement in the beam shapes compared to those shown in Figure 3. Also note that the same colour scales have been used as in Figure 3. This makes it easy to see that sidelobe levels in the optimised beam are much lower than that of the element beams.

In this paper, we present experimental results obtained with a prototype PAF. This prototype PAF has been installed in the focal plane of one of the WSRT 25-m dishes. Each Vivaldi receiving element has its own LNA. Due to the large physical size of the front-end system, these LNAs cannot be cryogenically cooled, leading to a higher system temperature compared to the current WSRT receivers. This, however, is largely offset by the higher aperture efficiency that PAFs offer. The system temperature of the first prototype PAF used by us was about 125 K. Recently, a new prototype with much better performance ($T_{sys}$ ~70 K) has been installed. This improvement in system temperature makes us confident that the requirement for the final system of $T_{sys}$ = 50-55 K will be met. The data from all PAF elements are, after down-conversion, recorded and stored in real-time. Much of the back-end of the prototype system consists of recycled hard- and software that earlier was used in the Lofar Initial Test System, while for the real-time beam forming we also use a Lofar beamformer. The fact that the real-time signals are stored means that all further experiments, such as beam forming, can be also done later off-line. It also means that the same observation can be used for several independent technical experiments. The same prototype PAF is now used at the GMRT for testing for this telescope. A close-up of this prototype is shown in Figure 1.

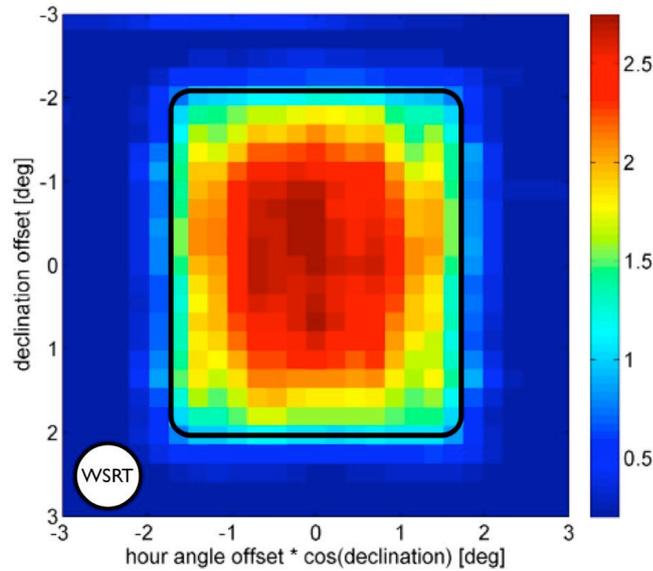

**Figure 5.** Sensitivity on the sky of the WSRT dish fitted with an PAF as measured with the first prototype PAF. The black line denotes the half-power points. For comparison, the field of view of the current WSRT system is also shown.

Figure 3 shows the beam patterns that were measured on the sky, in a single polarisation, for all elements of the PAF. Each panel covers the same 3° x 3° on the sky and shows the primary beam of a given element. It is clear that each element is sensitive to a slightly different area on the sky and that together they cover about a region of 8 deg$^2$. One can clearly see the large optical distortions in the reception patterns of the elements that are close to the edge of the PAF. However, because the PAF fully samples the radiation field in the focal plane, the signals from all elements can be combined to optimise the response of the telescope into a certain direction. In this way, the aperture efficiency of the telescope can be much increased and the shape of the combined beam can be controlled. Therefore, even for telescopes where one wants to have only one beam on the sky, PAFs are a very effective technology. Moreover, this optimisation can be done for many directions simultaneously so that many optimised beams can be formed to cover a large region on the sky. This is the real power of radio telescopes fitted with PAFs. Examples are given in Figure 4. This figure shows the optimised beams for two directions, as well as the weights given to each element in order to form these optimised beam. Note that in this figure the same colour scale is used as in Figure 3. This shows that the sidelobe level of the optimised beams is much lower than that of the response of an individual element. Figure 4 also shows that the shapes of the optimised beams are much better than those of the elements themselves.

Figure 5 shows the sensitivity (*A/T*) on the sky of the WSRT dish equipped with the first prototype PAF. The black line shows the location of the half-power points and shows that the field of view is larger than 8 deg$^2$. For comparison, the field of view of the current WSRT is also shown. The huge gain in survey speed offered by the PAF system is obvious.

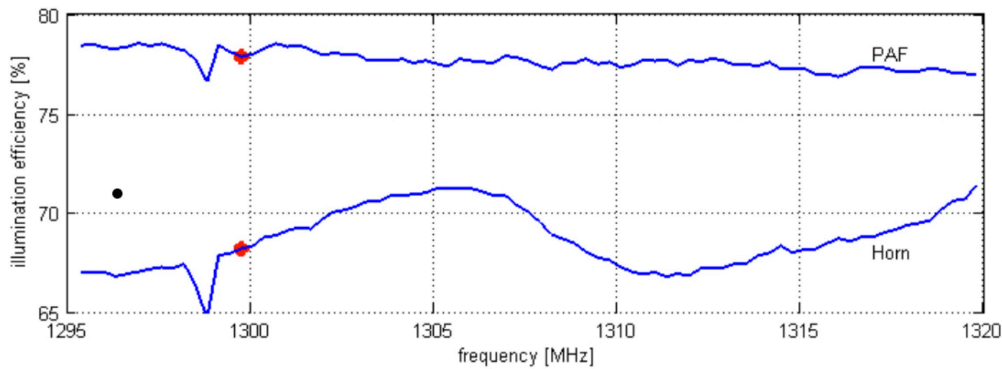

**Figure 6.** Illumination efficiency as function of frequency as measured for the current WSRT frontend (MFFE) and for the new Apertif prototype PAF installed in one of the WSRT dishes. The MFFE data clearly show the periodic modulation of *A/T* due to standing waves, while the Apertif PAF shows no sign of such effects.

Our most recent result is that a WSRT equipped with an PAF appears to suffer much less from effects due to standing waves in the dish. The current WSRT, as other radio telescopes, is known to suffer from such effects. Effectively, standing waves in the dish affect the illumination of the dish periodically as function of frequency. Due to this, the primary beam on the sky shows variations in size that are also periodic in frequency. This affects the observed spectrum of off-axis sources, in particular in those regions where the gradient in the primary beam is large. As a result, instead of being smooth, the spectrum of such sources show, sometimes strong, periodic variations. To make things worse, these variations are different in different polarisations. For the WSRT, the period of these effects is 17 MHz. This presents a huge complication for the calibration and analysis of wide-band spectral-line observations since this modulation of the spectra of off-axis sources has to be taken into account. However, measurements show that these negative effects are absent in the Apertif prototype. Figure 6 shows the illumination efficiency as function of frequency of the Apertif prototype as well as that of the current WSRT. The 17-MHz modulation of the illumination due to standing waves is clearly visible for the WSRT, while there is no sign of it in the Apertif data. Further study has shown that the absence of standing waves is due to a much lower reflection (a factor 1000!!) by the frontend of radiation back into the dish. The PAF absorbs more energy than the WSRT horn system, while for the small amount of radiation reflected back it moreover acts as a difusor. These results show that PAFs have excellent characteristics even in the case of a single-beam system.

## 4. First Light

First astronomical single-dish observations with the Apertif prototype were performed on a number of sources. An excellent demonstration of the power of the PAF system is shown in Figure 7. The right panel shows a 163 pointing mosaic of M 31 obtained by Braun using the existing full WSRT with a single feed [2]. The left panel shows a single pointing with the APERTIF prototype on a single telescope. A total of 11x11 optimised beams were formed with the Vivaldi signals of a single telescope pointing. The implication is that when all WSRT telescopes are equipped with a PAF system, the 163 pointing observation can be carried out in only one or a few pointings, demonstrating the impressive increase in survey speed.

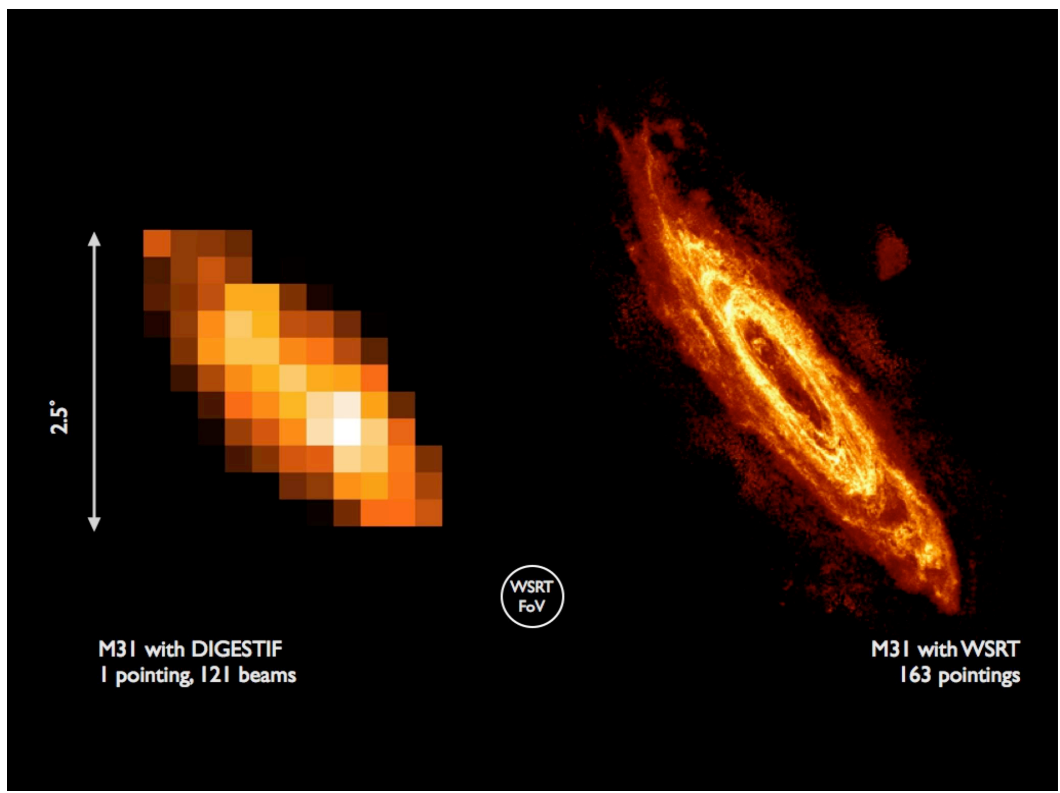

**Figure 7.** M31 observed with the WSRT. Left: A single-pointing measurement with the Apertif prototype, yielding the resolution of a single dish but covering the area in one observation. Right: A 163-pointing observation in interferometric mode, yielding the full resolution of the 3-km array. Once all WSRT antennas are equipped with an FPA, the same high resolution can be achieved over the full field of view of Apertif.

Further progress has been made by connecting the PAF equipped WSRT dish with 3 other WSRT dishes that still have the standard single-feed multi-frequency WSRT receiver (MFFE), and to use this to do interferometry. Figure 8 shows the fringes on 3C286 obtained by correlating the signal from an optimised PAF beam with that of a MFFE-equipped WSRT dish, as well as those between a single PAF element and an MFFE WSRT dish. For comparison, the fringes between two classic WSRT dishes is also shown. One can see that the amplitude of the fringes between optimised beam and the MFFE dish is about twice as large compared to those from the single PAF element correlated with the MFFE dish. This shows the effect of the beam optimisation. The difference in amplitude with the purely MFFE correlation is mainly due to the high system temperature of the PAF current prototype. This will improve when the new PAF prototype is installed.

Finally, in Figure 8 we show an image that was made doing full interferometry over 12 hours using correlations between the PAF prototype and three MFFE equipped WSRT dishes.

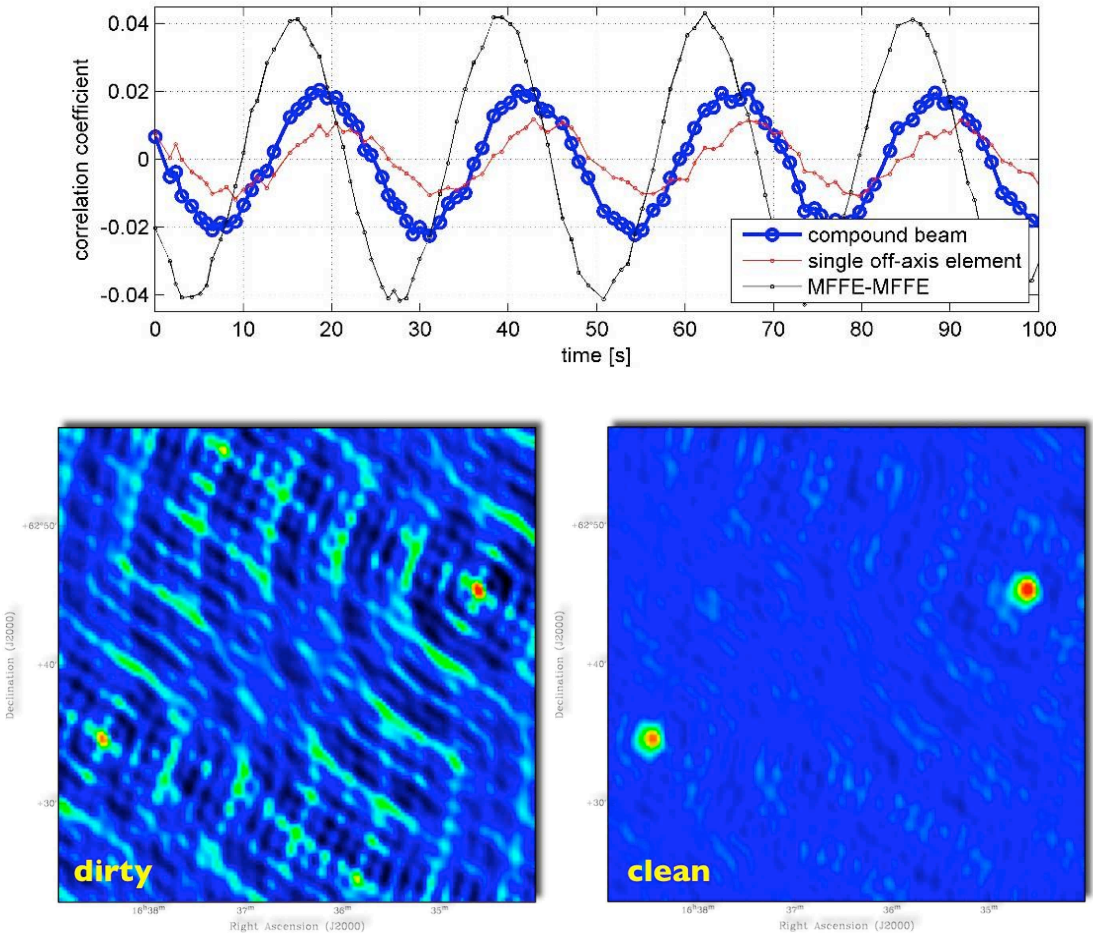

**Figure 8.** Fringes of 3C286 between an optimised beam of the FPA prototype and a MFFE WSRT dish, between a single FPA and the MFFE dish and between two MFFE equipped dishes. Bottom: First interferometric image made using a radio dish with an PAF. To make this image, the signal of the optimised beam of the PAF prototype was correlated with 3 WSRT dishes that used the single-feed MFFE.

The image shows the two sources 3C343 and 3C343.1 that are separated by about 20 arcminutes on the sky. As far as we know, this is the first interferometric image ever made using a radio dish fitted with a PAF.

## 5. Conclusions

We have briefly discussed possible science applications of Apertif, an upgrade to the WSRT that will increase its field of view with a factor 37 and double its observing bandwidth. The large, sensitive wide-field imaging surveys that will be done with Apertif will make it possible to address several important astronomical issues, such as the role of gas in the evolution of galaxies, the Galactic magnetic field, pulsars and the transient radio sky. We have shown the results obtained with prototype PAFs installed on one of the WSRT dishes. We have demonstrated that the field of view can be enlarged with a large factor (>30), we have also demonstrated that the high aperture efficiencies (75%) can be achieved with PAFs while the results also indicate that the system temperature of the final system will be close to 50 K. This system is also much more robust against the effects due to standing waves in radio dishes. All this holds great promise for the final Apertif system which will turn the WSRT into a very effective survey instrument.